%% file: main.tex
\documentclass[10pt,conference,table]{IEEEtran}
\usepackage{cite}
\usepackage{amsmath,amssymb,amsfonts}
\usepackage{algorithmic}
\usepackage{graphicx}
\usepackage{textcomp}
\usepackage{xcolor}
\usepackage{xspace}
\usepackage{fontawesome}
\usepackage{multirow}
\usepackage{rotating}
\usepackage{tikz}
\usepackage{soul}
\usepackage{makecell}
\usepackage{url}
\usepackage{enumitem}
\usepackage{booktabs}
\usepackage{fontawesome}
\usepackage{pifont}
\usepackage{amssymb} 
\usepackage[colorlinks=true, allcolors=black]{hyperref}
\usepackage{tcolorbox}
\usepackage[caption=false]{subfig}
\usepackage{lipsum}

\newcommand{\ie}{\emph{i.e.,}\xspace}
\newcommand{\eg}{\emph{e.g.,}\xspace}
\newcommand{\etc}{etc.\xspace}

\newcommand{\secref}[1]{Section~\ref{#1}\xspace}

\newcommand{\figref}[1]{Fig.~\ref{#1}\xspace}

\newboolean{showcomments}
\setboolean{showcomments}{true}

\ifthenelse{\boolean{showcomments}}
  {\newcommand{\nb}[2]{
    \fbox{\bfseries\sffamily\scriptsize#1}
    {\sf\small$\blacktriangleright$\textit{#2}$\blacktriangleleft$}
   }
  }
  {\newcommand{\nb}[2]{}
  }

\newcommand{\platform}{SEART Data Hub\xspace}
\newcommand{\demo}{\url{https://youtu.be/lCgQaA7CYWA}\xspace}
\newcommand{\website}{\url{https://seart-dh.si.usi.ch}\xspace}

\def\BibTeX{{\rm B\kern-.05em{\sc i\kern-.025em b}\kern-.08em
    T\kern-.1667em\lower.7ex\hbox{E}\kern-.125emX}}

\graphicspath{{figures/}}

\begin{document}

\title{
  \platform: Streamlining Large-Scale\\
  Source Code Mining and Pre-Processing
}

\author{
\IEEEauthorblockN{Ozren Dabi\'c, Rosalia Tufano, Gabriele Bavota}
\IEEEauthorblockA{\textit{SEART @ Software Institute - Universit\`a della Svizzera italiana}}
}

\maketitle

\begin{abstract}

Large-scale code datasets have acquired an increasingly central role in software engineering (SE) research. This is the result of (i) the success of the mining software repositories (MSR) community, that pushed the standards of empirical studies in SE; and (ii) the recent advent of deep learning (DL) in software engineering, with models trained and tested on large source code datasets. While there exist some ready-to-use datasets in the literature, researchers often need to build and pre-process their own dataset to meet specific requirements of the study/technique they are working on. This implies a substantial cost in terms of time and computational resources. In this work we present the \platform, a web application that allows to easily build and pre-process large-scale datasets featuring code mined from public GitHub repositories. Through a simple web interface, researchers can specify a set of \emph{mining criteria} (\eg only collect code from repositories having more than 100 contributors and more than 1,000 commits) as well as specific pre-processing steps they want to perform (\eg remove duplicates, test code, instances with syntax errors). After submitting the request, the user will receive an email with a download link for the required dataset within a few hours. A video showcasing the \platform is available at \demo.

\end{abstract}


\input{introduction}
\input{tool}
\input{action}

\input{future}
\input{conclusion}

\section*{Acknowledgment}
This project has received funding from the European Research Council (ERC) under the European Union's Horizon 2020 research and innovation programme (grant agreement No. 851720). 

\bibliography{bibliography}
\bibliographystyle{IEEEtran}

\end{document}

%% file: introduction.tex
\section{Introduction} \label{sec:intro}

Large-scale code datasets have been used in software engineering research to run empirical studies investigating specific characteristics of source code (\eg its naturalness \cite{Hindle:acm2016}), its evolution over time \cite{Behnamghader:emse2017}, how it is reused by developers \cite{Mojica:ieee2014}, \etc The recent born and wide success of deep learning (DL) for software engineering (DL4SE) as a research area has pushed even more the need for large-scale code datasets. To teach DL models how to deal with code-related tasks (\eg code generation \cite{Svyatkovskiy:fse2020}, code summarization \cite{Yao:ase2018}, code completion \cite{ciniselli2022}, test case generation \cite{Tufano:ast2022}), they need millions of examples showcasing the task (\eg pairs of $\langle$\emph{code}, \emph{textual\_summary}$\rangle$ to train for code summarization).

Despite the primary source for these datasets often being the same (\ie repositories on GitHub) and the availability of public code datasets (\eg CodeSearchNet \cite{husain2020}, Software Heritage \cite{cosmo2023}, The Stack v2 \cite{lozhkov2024}), researchers frequently tend to mine and pre-process their own dataset rather than reuse existing ones. This is usually due to specific requirements related to the research performed. For example, a study focusing on the production code may require the exclusion of test code, while the proposal of a DL-based code generator may benefit from the removal of code featuring syntax errors (to avoid showing the model wrong implementations). 

While one may just remove unwanted instances from existing datasets, the code archived in available datasets may soon become obsolete. For instance, one of the most used datasets in SE research is CodeSearchNet \cite{husain2020} which dates back to 2020. Such a lack of recently written code may be fine for some studies, but problematic for others. For example, empirical studies showed the \emph{concept drift problem} associated to coding assistants \cite{Ciniselli:icpc2024}, with code completion models decreasing their accuracy when new versions of a language are released, due to previously unseen coding constructs (thus implying the need for re-training on newer code).

To support researchers in building large-scale source code datasets, we present the
\platform, a web platform to continuously mine and process code from public GitHub repositories. The mining step is in charge of downloading the code and making sure it stays up-to-date with its  latest version available on GitHub. This means that (i) code files of already mined repositories will be updated when changes to them occur, with the files being deleted in case they are removed from the repository; and (ii) newly added files in already mined repositories will be collected, to keep the stored code aligned with that of the online repository. The processing step instead extracts information needed to automate specific pre-processing steps required by the user via a web interface. For example, we parse the code to check for syntax errors and to generate an AST-based representation of it. In this way, the user can customize the dataset they want to build via a handy GUI (\figref{fig:form}). Since the processing steps require language-specific parsers, our current implementation only supports Java and Python, but it is designed to be easily extended to other languages. \platform hosts, at the time of writing, over 22M Java and Python files mined from $\sim$316k repositories. On average, 1.3k new repositories are added every day.

\platform is an open source project \cite{dabic2024duatahub} and it is deployed at \website.

%% file: tool.tex
\section{\platform in a Nutshell} \label{sec:tool}

The \platform consists of four main components: (i) an intuitive \emph{web user interface}, that allows users to define the characteristics of the desired dataset; (ii) a \emph{back-end server}, which is responsible for managing and executing dataset construction requests; (iii) a \emph{crawler}, that collects and analyzes code from public GitHub repositories; and (iv) a central \emph{PostgreSQL database} for persisting data. We describe each component in the following subsections, with the last two components being described together.


\subsection{Web User Interface} \label{sec:user_interface}

\begin{figure}[ht!]
    \centering
    \includegraphics[width=\linewidth]{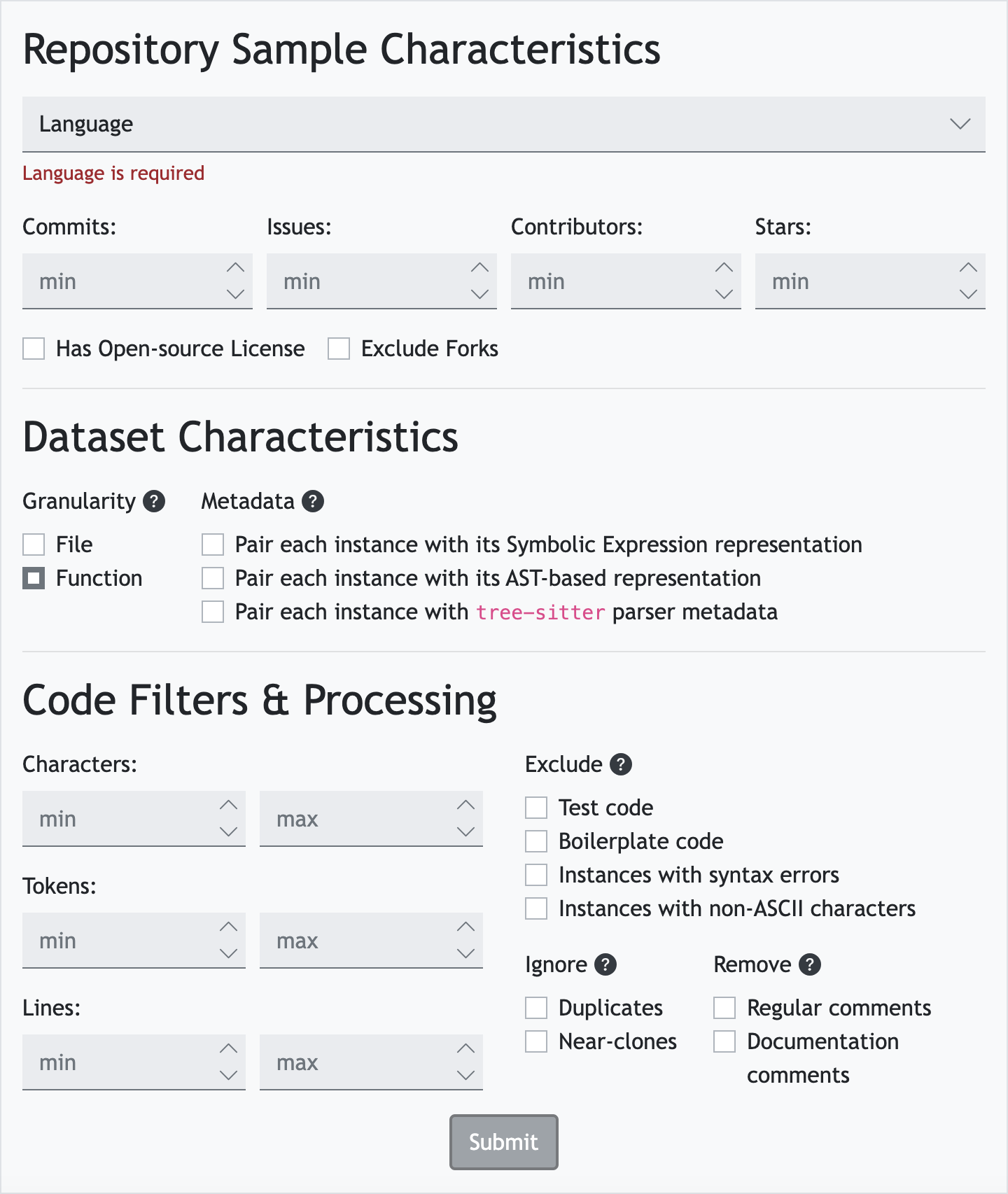} 
    \caption{Code dataset construction form.}
    \label{fig:form}
\end{figure}

The \emph{web user interface} is the only component with which the user (\ie a researcher interested in building a dataset) interacts with and it is designed to be intuitive and user-friendly. To get started, users must create an account by providing a valid email address that will be verified. Once logged into the platform, they can use the dataset construction form (\figref{fig:form}) to request the creation of a dataset. While the possible filters a user can exploit are described here, the way in which we extract the data needed to make them work is described in the subsections detailing the crawling, code analysis, and persistence mechanism of the \platform.

The form is organized into three main sections. The first section is the \emph{Repository Sample Characteristics}, featuring filters that are applied at the repository level (\ie to include only code from repositories that meet the specified requirements). The only mandatory filter to specify is the \emph{Language}, which in the current implementation can be either Java or Python, depending on the language of interest. Repositories can also be filtered by  quantitative metrics acting as proxy for their level of activity (\ie number of \emph{commits} and \emph{issues}), popularity (\ie number of \emph{stars}), and community size (\ie number of \emph{contributors}). Finally, users can decide to retrieve only projects explicitly accompanied by an open source license and to exclude forks. \smallskip

Next up is the \emph{Dataset Characteristics} section, in which the user can specify the dataset granularity, choosing between \emph{file} and \emph{function}. In the former case, the dataset will feature entire files as instances, while in the latter functions will be automatically extracted from each file. Other than that, the user can also specify which meta-data should be present in the dataset. This includes the \texttt{AST}, \texttt{symbolic\_expression} and \texttt{tree-sitter} parser metadata. \texttt{Tree-sitter} \cite{brunsfeld2024} is the tool we build upon for supporting source code analysis (details in \secref{subsec:analyzer}). The \texttt{AST} representation is provided in form of an Extensible Markup Language (XML) string, in which each tree node is an XML element and contains information about the node's type, as well as its start and end positions in the file. Such a representation can be useful both for empirical studies as well as starting point for more expressive code representations for DL models (similar to what is done in \cite{Yang:DSN2021}). Since the \texttt{AST} representation significantly increases the dataset size, the \texttt{symbolic\_expression} \cite{enwiki:1213363014} is an alternative. It provides a more compact and human-readable representation, featuring the same data found in the \texttt{AST}, save for the positional information which is not present. Finally, the \texttt{tree-sitter} parser metadata includes the semantic version tag of the tree-sitter binding release which was used to parse each instance. Given that the binding and its grammars change over time, this information can be useful for reproducibility purposes. \smallskip

The last part of the form is the \emph{Code Filters \& Processing} section, which allows the user to select/exclude code instances having specific characteristics. 

In particular, the user can (i) select code instances (\ie files or functions, depending on the selected granularity) meeting specific size requirements in terms of \texttt{characters}, \texttt{tokens}, and/or \texttt{lines} (\eg only functions having more than 5 but less than 20 lines); (ii) exclude irrelevant instances such as \emph{test code}, \emph{instances with syntax errors}, \emph{instances with non-ASCII characters}, and \emph{boilerplate code}\footnote{Only applicable for datasets at function-level granularity.} (\eg \texttt{getters}, \texttt{setters}, \texttt{toString} functions); (iii) exclude \emph{exact duplicates} or \emph{near-clones} (\ie same \texttt{AST}, but potentially different in terms of identifiers/literals) from the dataset; and (iv) pre-process each instance to remove \emph{regular comments} (\ie inline/block comments within the code instance) and/or \emph{documentation comments} (JavaDoc for Java, docstring for Python).

\begin{figure}[ht!]
    \centering
    \includegraphics[width=\linewidth]{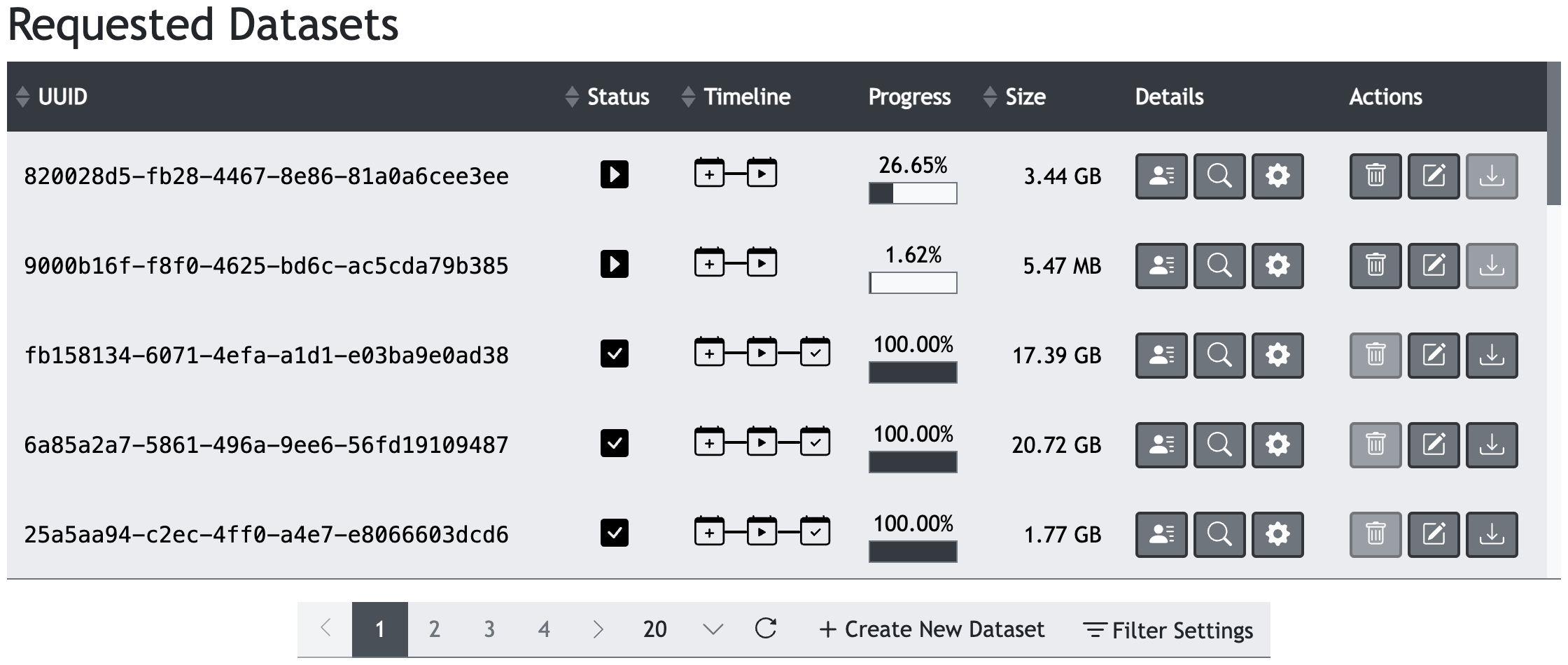} 
    \caption{Submitted requests dashboard}
    \label{fig:dashboard}
\end{figure}

Besides the dataset construction form, the user has access to a dashboard page (\figref{fig:dashboard}), allowing them to monitor in real time the status of their dataset construction requests, cancel them, and download the related dataset once a request has been processed. 

\subsection{Back-end Server}

The \emph{back-end server} is responsible for managing and executing dataset construction requests submitted through the web user interface. 
User requests arrive in JavaScript Object Notation (JSON) format and are stored in the PostgreSQL database to ensure their persistence. The request is then added to a task execution queue, where it awaits execution. 

A task scheduler is responsible for managing this queue, processing tasks in a first in first out (FIFO) manner. When selecting a task for execution, the scheduler creates a new task executor thread, which is responsible for constructing the dataset. A fixed number of task executors can be active at any given moment, and this number can be modified at runtime by the administrator in a specific setting panel. The executor retrieves the task from the queue and starts the process. 

The first step involves transforming the dataset specification into two separate queries: one which opens a direct read stream to the database, and another which estimates the number of instances that will be exported. The latter is needed to provide users with a real-time feedback about the execution of a request, as shown in the progress bar in \figref{fig:dashboard}. To maximize the performance, these queries are executed in parallel. Once the executor starts receiving content from the database (\ie either code files or functions, depending on the selected granularity), it performs the pre-processing operations required by the user as described in \secref{subsec:analyzer} (\eg removal of \emph{regular comments}).

Each matched and processed instance is then converted into JSON strings, before being compressed and written directly to the output file, which is in JSON Lines (JSONL) format. The executor continues this process until all matched instances have been written on file. This process can take hours to complete, which is why the executors are also in charge of regularly updating their progress and informing the user of the task's completion via email.



\subsection{Crawling, Analyzing, and Storing Code} \label{subsec:analyzer}

The \emph{Crawler} is responsible for collecting and analyzing source code from GitHub repositories, as well as making sure that already collected code is always up-to-date. The \platform is set to mine all public repositories (featuring at least one Java or Python file) indexed by GitHub Search \cite{dabic:msr2021}, an online platform we built which continuously mines metadata about GitHub repositories having at least ten stars. Note that GitHub Search does not provide the user with the possibility of downloading the repositories' code, but only to identify GitHub repositories matching specific requirements (\eg having at least 100 commits). 

The choice of relying on GitHub Search \cite{dabic:msr2021} provides ``\emph{for free}'' an always updated list of projects to mine, without the need of interacting with the GitHub APIs. 

The drawback is the exclusion of projects having less than ten stars. However, (i) given the amount of code available on GitHub, our platform will still offer very large-scale datasets; and (ii) a project not included in the first place as having less than ten stars, will be included as soon as it crosses this threshold. Once the list of repositories is received from GitHub Search \cite{dabic:msr2021}, we process them sequentially ordered by the date of their last commit (from the oldest to the newest). We store in our PostgreSQL database the repository owner and name as it appears in GitHub. This will serve as its primary key. We also retrieve and store the repository metadata needed for the top part of the dataset construction form (\ie \emph{Repository Sample Characteristics} --- see \figref{fig:form})  from GitHub Search. In particular, we store for each repository: its license (if any),  a boolean flag indicating whether it is a fork or not, and its number of  commits, issues, contributors and stars. 

A shallow clone of the repository is then performed (\ie only its latest snapshot is downloaded), and we store its \texttt{last\_commit} date and \texttt{last\_commit\_sha}. This information will be later used to keep the repository's code updated over time by checking whether new commits have been performed since the time we processed it. Each code file written in Java or Python is then processed by relying on \texttt{tree-sitter}\cite{brunsfeld2024}, a parser generator tool and incremental parsing library originally developed by GitHub for the Atom text editor. We chose to use \texttt{tree-sitter} due to its  error recovery capabilities, plug-and-play language support and unified syntax-tree navigation for all languages \cite{tsdocs}. This allows us to use a single parsing framework for virtually all programming languages, and any analysis algorithms and heuristics can be easily shared across different languages. This simplifies the future extensibility of our tool to additional programming languages. Furthermore, its incremental nature allows to parse large files without running into memory issues, while error recovery mechanisms ensures that we can still extract useful information from files with syntax errors. 

Since \texttt{tree-sitter} is written in C and most of our platform infrastructure runs on the Java Virtual Machine, we had to develop a way to interact with the parser APIs from the Java programming environment. This resulted in a custom API binding which uses the Java Native Interface (JNI) to communicate with the lower-level C APIs. This binding is itself a standalone project that we have open-sourced \cite{dabic2023java}.

Using such a project, we implemented code analysis and transformation tools needed to extract the information required by the filters shown in the dataset construction form (\figref{fig:form}). The code analysis tools traverse the syntax tree structure generated by \texttt{tree-sitter}, check for syntax errors, and compute various metrics, such as the number of code tokens, characters, lines, as well as an incremental hash over the tree's nodes. Code analysis also allows to identify exact duplicates (\ie code instances having the exact same hash when only considering code nodes) and near-clones (\ie code instances having the same \texttt{AST} when only considering code nodes). 

\eject

Finally, such a step is responsible for storing the additional metadata the user can collect in the \emph{Dataset Characteristics} part of the dataset construction form (\ie \texttt{AST} representation, \texttt{symbolic expression}, and \texttt{tree-sitter} parser metadata). Transformation tools are instead used to modify the syntax tree and its corresponding source code. This involves removing or replacing nodes in the tree, or even creating new ones. The only scenario for which we currently use transformation tools is the removal of comments from the instances when required by the user. 

All code analysis steps are performed at mining time (\ie while a project is crawled), while the transformation tools are only run when a request by the user is submitted and requires specific code processing (in our case, the removal of comments). Indeed, code is stored in our database as is, including all its comments.

Finally, a set of language-specific heuristics is applied to identify \emph{test} and \emph{boilerplate code}. For example, we flag test code by looking for the path in which files are contained (\eg ``\texttt{/test/}'') and by relying on naming conventions (\eg the file name contains ``\texttt{test}''). Similarly, for boilerplate code we implemented heuristics aimed at identifying getters, setters, constructors, \etc The interested reader can inspect the implemented heuristics in our GitHub repository --- see FAQ section in the README \cite{dabic2024duatahub}.

All above-mentioned information is stored both at file and function-level. While more space-consuming, such an approach significantly reduces the time users will wait for the dataset creation, since everything is pre-computed at both granularity levels we support.

To improve the efficiency of the overall process, we analyze the content of each repository in parallel (\ie we allocate different threads to the analysis of different files). No more than eight threads are allocated in parallel in the current settings we use. 

\subsubsection{Keeping Code Up-to-date} While we have outlined the process of collecting new data, keeping the existing data up-to-date is equally important. Once completed a first pass on all repositories returned by GitHub Search, we query again this service to retrieve (i) newly created repositories, and (ii) repositories which have been updated since we analyzed them (relying on the \texttt{last\_commit} info we stored). Newly created repositories are analyzed as previously described. For the updated ones, instead, we shallow clone them from the last recorded commit date in our database (\ie we download the repository history from the last commit we analyzed up to the most recent commit). Then, we compute the git \texttt{diff} between the snapshot we analyzed and the most recent one, obtaining the list of all changed/deleted/added files. Actions performed vary depending on the type of change. Deleted files are simply removed from the database, while all added or modified files are analyzed and stored or updated, respectively.

\eject


%% file: action.tex

\begin{figure}[htb!]
    \centering
    \includegraphics[width=\linewidth]{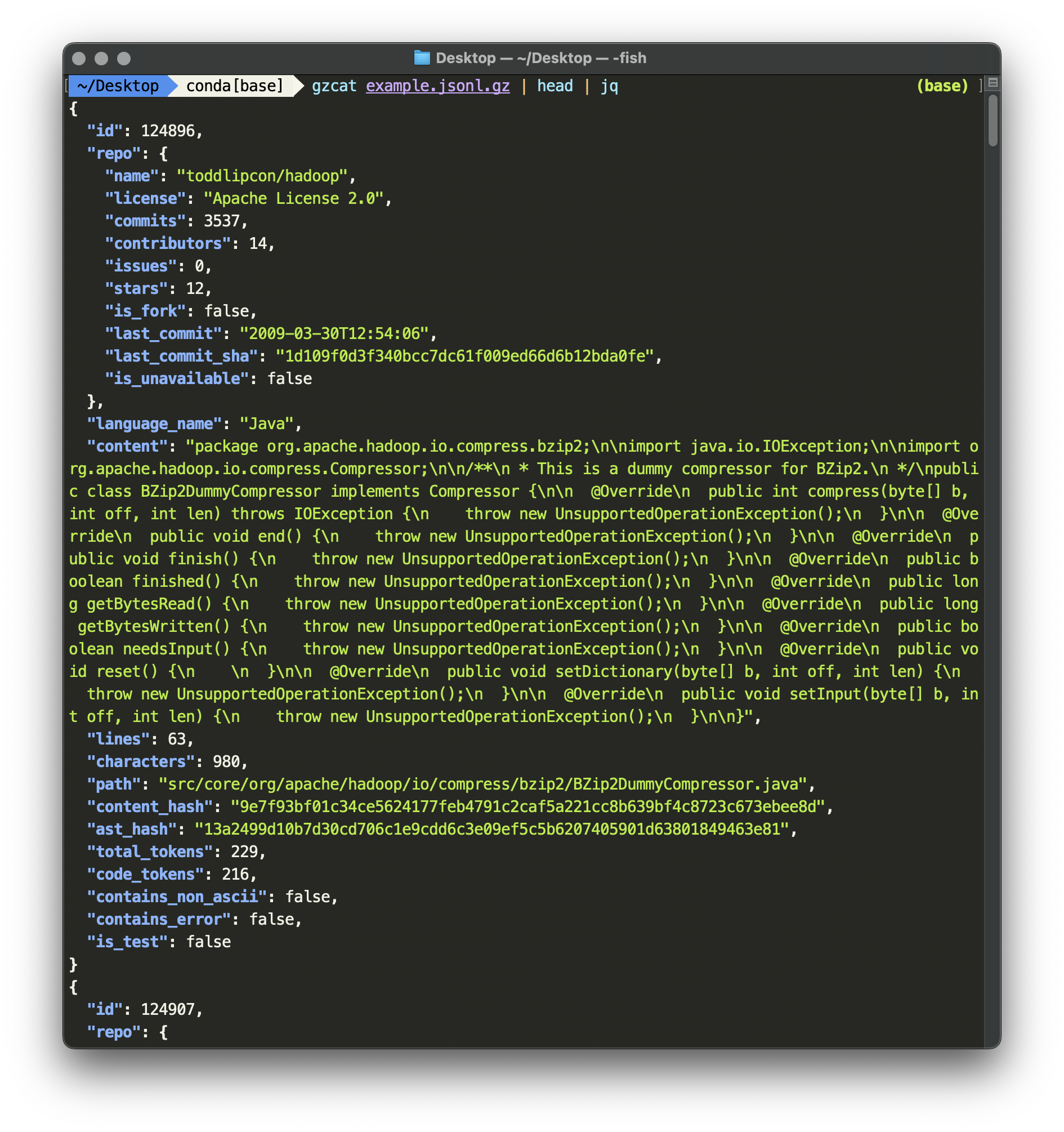} 
    \caption{Pretty-printed segment of the dataset.}
    \label{fig:result}
\end{figure}

\section{\platform in Action} \label{sec:action}

We now illustrate a typical dataset construction process. Let us assume that we are interested in building a dataset featuring non-test Java files with at least five lines of code, coming from non-fork projects and without duplicates. After submitting such a request via the dataset construction form, the entire process starts. With the current data available in the \platform, it would take $\sim$8 hours to complete. The resulting dataset contains $\sim$6.5M files, and its uncompressed size is roughly 11.7 GB of source code and associated metadata. \figref{fig:result} shows an excerpt of the created dataset.

\begin{figure}[htb!]
    \centering
    \includegraphics[width=\linewidth]{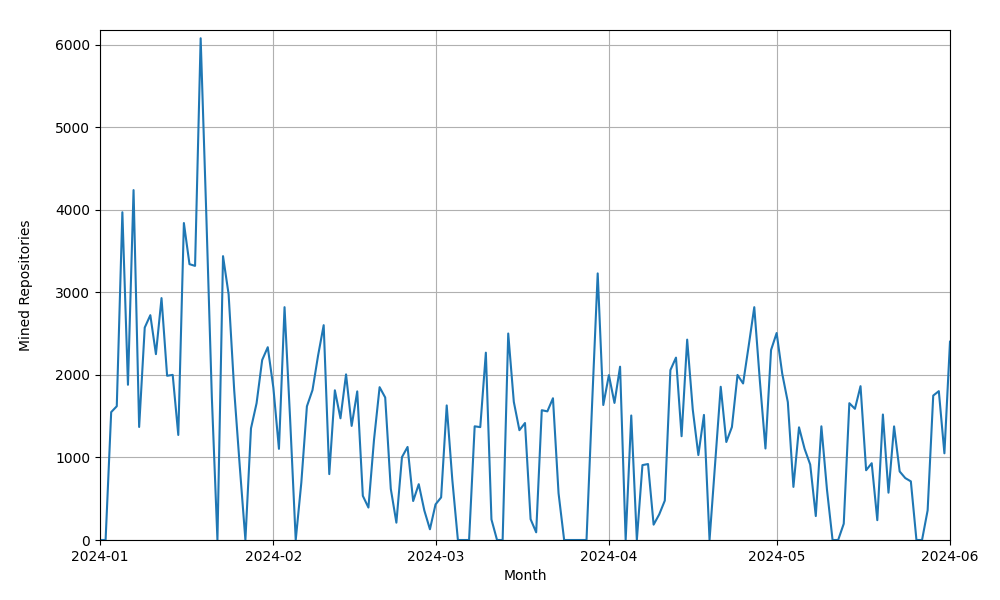} 
    \caption{Daily number of repositories mined from January to June 2024.}
    \label{fig:mining}
\end{figure}

Since its deployment in September 2023, \platform has mined code from more
than 316k repositories. From these repositories, it has analyzed more
than 22M files, featuring over 202M functions. This equates to 130 GB of source code mined, and more than 3B lines of code analyzed. As for the actual mining rate, the amount
of data we mine in a single day can range anywhere from 10 to 6,000 repositories. This difference can be attributed to the fact that repositories differ in the number of files they feature, with files also varying in terms of size and complexity. Overall, the number of repositories mined per
day averages around 1.3k. \figref{fig:mining} shows the number of
repositories mined each day within the first half of 2024, as extracted from the
crawler logs. Note that there were some days in which the crawler did not mine
any repositories, which can be attributed to either our servers being down due
to issues, or the crawler itself being down for maintenance. 

As previously explained, we mine repositories in chronological order (starting from the one updated least recently). 

As of Today, we have mined all repositories which have been last updated in the beginning of 2023. Looking at the list of repositories provided by GitHub Search and at our current mining rate, we estimate that we will have mined all repositories featuring Java/Python files by the end of 2024.

%% file: future.tex
\section{Future work} \label{sec:future}

We plan to extend the \platform in several directions. First, we plan to introduce support for other programming languages, such
as C, C++, JavaScript and TypeScript. Second, we are designing an opt-out mechanism for repositories that do not what their code to be stored in our database. Third, we intend to collect feedback from the research community about the functionalities offered by our platform, introducing support for
additional code metrics and transformation operations as requested. Fourth, we will
also continue to improve the performance and scalability aspects of the
platform, ensuring that it can handle the ever-growing amount of data. Finally, we plan to introduce support for additional
dataset types. Although we focused on collecting data for models that work with
source code, DL models have been used in various software engineering
tasks, such as the automation of code reviews
\cite{tufano2021}. These tasks require specific types of information (\eg reviewers' comments) which are not currently part of our platform, and that we plan to add.

%% file: conclusion.tex
\section{Conclusion} \label{sec:conclusion}

We presented the \platform, a platform for collecting and pre-processing source code
from public GitHub repositories with the goal of helping researchers create large-scale
code datasets for conducting empirical studies and training/evaluating DL models. The
platform is deployed at \website and available as an open source project hosted on GitHub \cite{dabic2024duatahub}.
We welcome contributions in the form of pull requests, issue reports and feature discussions. 
